\documentclass[conference]{IEEEtran}
\IEEEoverridecommandlockouts

\usepackage{cite}
\usepackage{amsmath,amssymb,amsfonts}
\usepackage{algorithmic}
\usepackage{graphicx}
\usepackage{textcomp}
\usepackage{xcolor}
\usepackage{amsmath}

\usepackage{mathtools} 
\usepackage{amsfonts}
\usepackage{amssymb}
\usepackage{accents}

\usepackage{algorithm}
\usepackage{algorithmic}
\usepackage{multirow}

\def\BibTeX{{\rm B\kern-.05em{\sc i\kern-.025em b}\kern-.08em
    T\kern-.1667em\lower.7ex\hbox{E}\kern-.125emX}}
    
\newlength\myindent
\begin{document}

\title{\huge{PPO-ABR: Proximal Policy Optimization based Deep Reinforcement
Learning for Adaptive BitRate streaming}}


\author{\IEEEauthorblockN{Mandan Naresh, Paresh Saxena and Manik Gupta}
\IEEEauthorblockA{\textit{Dept. of CSIS, BITS Pilani}\\
Hyderabad, India \\
\{p20180420, psaxena, manik\}@hyderabad.bits-pilani.ac.in}

}

\maketitle

\begin{abstract}


Providing a high Quality of Experience (QoE) for video streaming in 5G and beyond 5G (B5G) networks is challenging due to the dynamic nature of the underlying network conditions. Several Adaptive Bit Rate (ABR) algorithms have been developed to improve QoE, but most of them are designed based on fixed rules and unsuitable for a wide range of network conditions. Recently, Deep Reinforcement Learning (DRL) based Asynchronous Advantage Actor-Critic (A3C) methods have recently demonstrated promise in their ability to generalise to diverse network conditions, but they still have limitations. One specific issue with A3C methods is the lag between each actor's behavior policy and central learner's target policy. Consequently, suboptimal updates emerge when the behavior and target policies become out of synchronization. 
In this paper, we address the problems faced by vanilla-A3C by integrating the on-policy-based multi-agent DRL method into the existing video streaming framework. Specifically, we propose a novel system for ABR generation - Proximal Policy Optimization-based DRL for Adaptive Bit Rate streaming (PPO-ABR). Our proposed method improves the overall video QoE by maximizing sample efficiency using a clipped probability ratio between the new and the old policies on multiple epochs of minibatch updates. The experiments on real network traces demonstrate that PPO-ABR outperforms state-of-the-art methods for different QoE variants.


\end{abstract}


\section{Introduction}\label{section:introduction}
Due to the widespread use of the Internet, the volume of multimedia traffic has increased, including video streaming. 
The Cisco annual Internet Report projects that by 2023, 69\% of the world's population will have access to the Internet, with Internet video traffic significantly outnumbering other Internet traffic. 
In order to ensure seamless video streaming, Dynamic Adaptive Streaming over HTTP (DASH) \cite{dash} uses an adaptive bit rate (ABR) algorithm to send the video encoded at a specific bitrate based on the network conditions. Several ABR algorithms such as 
RB \cite{rb}, BB \cite{bb}, BOLA \cite{bola}, and Robust-MPC \cite{mpc} use network conditions including throughput estimation, playback buffer occupancy or a combination of both for bitrate estimation with the aim to enhance the QoE for end users. 
However, traditional ABR algorithms are designed with specific network conditions and traffic pattern assumptions. As a result, they may not perform optimally in networks where network conditions and traffic patterns are subject to rapid and unpredictable change. Recently, several data-driven deep reinforcement learning (DRL) approaches, including Pensieve \cite{pensieve}, $A^{2}BR$ \cite{refa}, VSiM \cite{refb}, NANCY \cite{nancy}, AL-FFEA3C \cite{deeprl}, AL-AvgA3C \cite{deeprl}, MARL-A3C \cite{marl}, SAC-ABR \cite{sac-abr} and ALISA \cite{alisa} are proposed to improve the ABR algorithms. 
DRL is a branch of deep learning that deals with how agents should behave depending on the state of the environment. In DRL, a policy is created to maximize the expected cumulative reward. The policy is the mapping function from states of the environment to actions. Pensieve \cite{pensieve}, being one of the first DRL-based methods for ABR generation, is built upon the basic vanilla-A3C algorithm, whereas ALISA \cite{alisa}, being the latest DRL-based ABR method, utilizes soft updates with an A3C algorithm. Both Pensieve and ALISA update the
ABR control policy based on the current network conditions and past decisions, and it is able to
identify policies that outperform traditional ABR algorithms. 

However, these state-of-the-art DRL-based methods suffer from two key drawbacks: (i) there is a lag between each actor's behavior policy and the central learner's target policy. Consequently,
suboptimal updates emerge when the behavior and target policies
become out of synchronization, and (ii) there is a constraint on the divergence between the new and the old policies. Due to these constraints, these algorithms may result in imprecise throughput prediction when there are fluctuations in the network, re-buffering at the client's device, and  inaccurate bitrate selection impacting the overall QoE for the end users. 
 To resolve the above issues, we propose the integration of Proximal Policy Optimization-based DRL for ABR (PPO-ABR) to use a clipped probability ratio for constraining the divergence between the new and the old policy parameters. Our experimental results show that PPO-ABR improves overall video QoE as compared to other state-of-the-art methods. 
 
 The rest of the paper is organized as follows:
 Section
\ref{section:Background} presents the relevant background on reinforcement learning and on-policy RL methods.  Section \ref{approach} presents the design of the proposed PPO-ABR algorithm. We present the experimental
setup and results in Section \ref{section:Experimental details} where we include both training and testing results. Finally, we conclude our work in Section \ref{section:Conclusion}.
\section{Background}\label{section:Background}



RL \cite{suttonbarto} is a learning process that is adaptive to dynamic environments, even in cases where there is little or no prior information. By learning from its mistakes, an agent seeks to optimize its long-term return in the future. The agent's interactions with the environment are described using a Markov decision process (MDP), where at each time step (represented by $t = 0,1,2,3,...$), the agent is situated in a specific state ($s_t$), chooses an action from a set of available actions ($a_t \in A$), and then receives a reward ($r_t = R(s_t, a_t)$) based on its action. The goal of the agent is to find a policy $\pi(s,a)$ that maps states to actions. The state-value function is given by $V^\pi_\phi (s) = E\Bigg[\sum_{k=0}^{\infty} \gamma^k r_{t+k}  | s_t =s,\pi_\phi \Bigg]$
and the action-value function is given by, $ Q^\pi_\phi (s,a) = E\Bigg[\sum_{k=0}^{\infty} \gamma^k r_{t+k} | s_t =s, a_t = a, \pi_\phi \Bigg]$
where, $\gamma \in [0,1)$ is a discount factor. The basic on-policy RL method is a vanilla policy gradient method \cite{a3c} where policy parameters are updated after the calculation of the total reward at the end of the episode instead of a single-step. The policy gradient is given by, 
\begin{equation}
\label{eq11}
\nabla_{\phi_{k}}= \sum_{t=0}^{T} \nabla_\phi \log \, \pi_\phi(a_t,s_t)|_{{\phi_{k}}} A_\phi(s,a)
\end{equation}

where $A_\phi(s,a) = Q^\pi_\phi (s,a) - V^\pi_\phi (s)$ is the advantage function,   $\nabla_\phi$ is the policy optimization using a gradient operator, $T$ is the number of steps in the episode and $\phi_k$ is the  current policy parameters. However, the vanilla policy gradient suffers from high variance and high training time due to value estimates being calculated at the end of the episodes instead of every time step. To address these issues, actor-critic methods \cite{a3c} are proposed. 
These methods have two components: an actor represented by a policy $\pi$ and a critic represented by an estimate of the action-value function. Neural network function approximators are typically used to represent both of them. With parameters $\theta$, the critic estimates the current policy’s value function. The main goal of this method is to reduce the variance using single-step state-value estimates. The single-step state-value estimates are derived using a temporal difference $(\delta)$, and it is given by:

\begin{equation}
\label{eq22}
\delta = V^\pi_\phi (s_t) + \gamma V^\pi_\phi (s_{t+1},\phi)- V^\pi_\phi(s_{t},\phi)
\end{equation}


The gradient operator $\nabla$ is used to define the policy and critic updates with regard to its parameters $\phi$ and $\theta$, respectively:

\begin{equation}
\label{eq33}
\Delta\phi = \phi + \alpha_p \delta \nabla \pi_\phi(s_{t+1},a_{t+1},\phi)
\end{equation} 

\begin{equation}
\label{eq34}
\Delta\theta = \theta + \alpha_c \delta \nabla V^\pi_\phi(s_t,\theta)
\end{equation}

where $\alpha_p$ and $\alpha_c$ are the actor and critic learning rates, respectively. Furthermore, as an improvement, vanilla-A3C \cite{a3c} is proposed that uses several copies of the same agent with asynchronous updates. It is more efficient than the actor-critic methods because samples for data can be parallelized using several copies of the same agent resulting in an even smaller training time. In the vanilla-A3C algorithm, the current policy parameters ($\phi_{new}$) are updated based on previously collected experience with old policy parameters ($\phi$) after every $\kappa$ steps, i.e., after every $\kappa$ state-action pairs. The equation below represents the value function update for vanilla-A3C is:

\begin{equation}
\label{eq50}
 \resizebox{1.0\hsize}{!}{$%
\text{maximize}_\phi  \ V^\pi_{\phi_{new}} (s) = \kappa \nabla V^\pi_\phi(s)  + \kappa \sum_{s} \rho_{\pi_{\phi}}(s)  \sum_{a} \pi_{\phi_{new}}(a|s)A_\phi(s,a)
        $%
}%
\end{equation}


where $\rho_\pi(s)$ presents distribution of state-action pairs, $\pi_{\phi}$ represents the old policy and $\pi_{\phi_{new}}$ represents current policy. Note that $\sum_{a} \pi_{\phi_{new}}(a|s)A_\phi(s,a) \ge$0 aims to increase the value function, however,  $\sum_{a} \pi_{\phi_{new}}(a|s)A_\phi(s,a)<0$ can result in a decrease in the value function and in a increase of divergence between the old and the new policies. 

To alleviate this issue, the on-policy trust region policy optimization (TRPO) \cite{trpo} proposes Kullback–Leibler (KL) divergence  constraint to update the value function. The equation (\ref{eq50}) is rewritten with KL divergence constraint as follows:

\begin{equation}
\label{eq51}
 \resizebox{1.0\hsize}{!}{$%
   \begin{aligned}
    \text{maximize}_\phi  \ V^\pi_{\phi_{new}} (s)  = \kappa \nabla V^\pi_\phi (s) + \kappa E_{s \sim \rho_{\pi_{\phi}}, a\sim \pi_\phi}\Bigg[r(\phi)\,A_\phi(s,a)\Bigg] \\
    \text{subject to} \ D_{KL}( {\pi_{\phi_\text{new}}}|| \pi_{\phi}) \leq \lambda
    \end{aligned}
            $%
}%
\end{equation}

where $r(\phi)= \frac{\pi_{\phi_{new}}(s,a)}{\pi_{\phi}(s,a)}$ is the importance sampling ratio, $D_{KL}( {\pi_{\phi_\text{new}}}|| \pi_{\phi}) = \sum_{a} \pi_{\phi_{new}}(s,a) \log \Bigg( 
\frac{\pi_{\phi_{new}}(s,a)}{\pi_{\phi}(s,a)}\Bigg)$ and $D_{KL}( {\pi_{\phi_\text{new}}}|| \pi_{\phi}) \leq \lambda$ is used to constrain the divergence between the new and old policies with $\lambda$ as a KL-divergence limit, $\lambda \in (0,1]$. We can rewrite equation (\ref{eq51}) to maximize only the second part, also known as the surrogate advantage objective, as follows:
\begin{equation}
\label{eq:eqtrpoobj}
\begin{gathered}
    \text{maximize}_\phi \ \kappa E_{s \sim \rho_{\pi_{\phi}}, a\sim \pi_{\phi}}\Bigg[r(\phi) \, A_\phi(s,a)\Bigg] \\
    \text{subject to} \ D_{KL}( {\pi_{\phi_\text{new}}}|| \pi_{\phi}) \leq \lambda
\end{gathered}
\end{equation}

Although TRPO provides constraints on the divergence between the new and the old policies, it can still lead to instability in policy updates. To address this issue, the on-policy PPO algorithm \cite{ppo} is proposed that uses a clipped probability ratio to constrain the divergence between the old and the new policy parameters. The objective function in PPO is derived from Equation (\ref{eq:eqtrpoobj}), and the maximization problem is given as:

\begin{equation}
\label{eq52}
 \resizebox{1.0\hsize}{!}{$%
   \begin{aligned}
  \text{maximize}_\phi  
    L^{clip}(\phi_{new}) = \kappa E_t\Bigg[\min\Bigg( L^{CPI}(\phi), clip(r(\phi) \,,1-\epsilon, 1+\epsilon)A_\phi(s,a)\Bigg)\Bigg] \\
    \text{subject to} \ D_{KL}( {\pi_{\phi_\text{new}}}|| \pi_{\phi}) \leq \lambda
    \end{aligned}
            $%
}%
\end{equation}

where $\epsilon$ is the hyperparameter for clipping and  $L^{CPI}(\phi) = \kappa E_t\Bigg[r(\phi) \, A_\phi(s,a)\Bigg]$ where CPI refers to a conservative policy iteration. From Equation (\ref{eq52}), the first term represents the TRPO unclipped surrogate objective, and the second term represents a modification of the TRPO surrogate objective 
using a clipped probability ratio $\epsilon$, which ensures that the $r(\phi)$ remains within the range $[1-\epsilon, 1+\epsilon]$. The PPO maximization considers the minimum of the clipped and unclipped objectives resulting in a smaller divergence between the new and the old policy parameters.

\section{Proposed on-policy ABR Method: PPO-ABR}
\label{approach}
\begin{figure}
\centering
\includegraphics[width=0.8\linewidth]{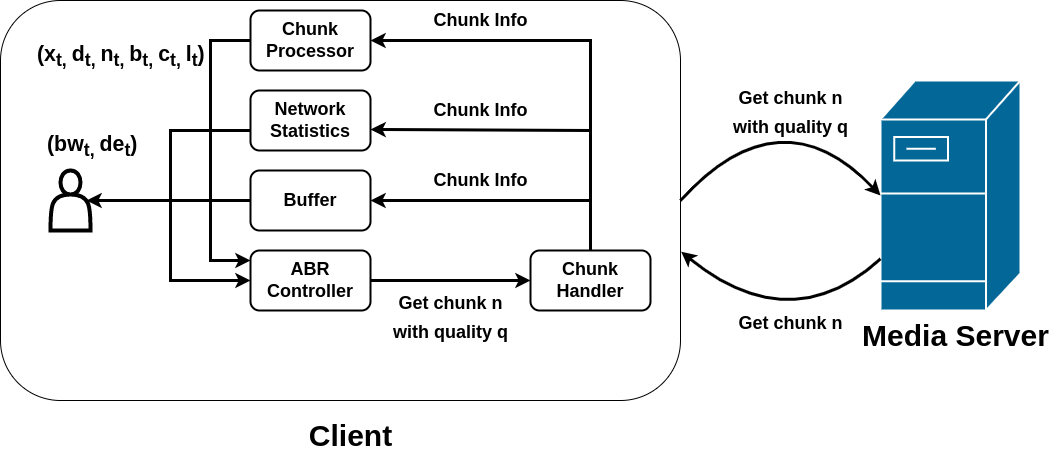}
\caption{System Model depicting multimedia streaming.}
\label{fig:SYSTEM_DESIGN}
\end{figure}

In this paper, we focus on the HTTP-based video distribution system, as shown in 
Figure \ref{fig:SYSTEM_DESIGN} that utilize the DASH framework for multimedia streaming.  
In such systems, the videos are stored on the server in separate chunks, where each chunk is encoded with a specific bitrate. The client then requests each chunk with the appropriate bitrate from the server using an ABR algorithm, where the ABR algorithm generates the bit rate based on factors such as the available network conditions and the capabilities of the client device. Specifically, an ABR algorithm selects the bitrate for each video chunk based on chunk processor input observations, including the number of chunks $(c_t)$, chunk
size $(n_t)$, chunk bitrate  $(l_t)$, size of the buffer $(b_t)$, throughput $(x_t)$, and download time $(d_t)$. Additionally, the ABR controller takes the network statistics such as bandwidth $(bw_t)$ and delay $(de_t)$ into account. 
    
For the state-of-the-art vanilla-A3C, the ABR controller uses multi-agent training with multiple actor and critic neural networks. Each agent is trained in parallel with its own environment based on several state inputs $ s_t = (x_t, d_t, n_t, b_t, c_t, l_t, bw_t, de_t)$. Moreover, each agent is trained and sends the local gradients to the central agent. Once the central agent has collected experience from the local agents, it updates its model parameters. Further, the central agent will make the decision to play the chunk with a specified bitrate to the chunk handler. The chunk handler sends the information about the chunk to the buffer and finally, the client will play the chunk $n$  with quality $q$ based on buffer occupancy.

In addition to being less sample efficient, the vanilla-A3C also has a high divergence between the target policy of the central
learner and every actor’s behavior policy. The suboptimal updates emerge when the behavior and target policies become out of synchronization. To address these issues, PPO-ABR uses a clipped probability ratio to constrain the KL-divergence between the new and the old policy parameters among several epochs instead of a single epoch as in vanilla-A3C.

Algorithm \ref{alg:2} presents the PPO-ABR algorithm and outlines the critical steps. The input to the algorithm is video samples, including hyperparameter setting for actor and critic networks and state input as $s_t = (x_t, d_t, n_t, b_t, c_t, l_t, bw_t, de_t)$.  The first step is dividing a video file into chunks. Each chunk is played at a specified bitrate using the selection of the action based on the current state and the policy and to store the corresponding reward at Line 12. The actor-network finds the policy $\pi_\phi(.|s_t)$, and the critic network estimates the state value function. The second step of this algorithm is to compute the advantage function using a current policy at Line 15. The third step is to compute the policy divergence between the new and the old policies using an important sampling ratio $(r(\phi))$ at Line 17. The fourth step is to update the actor parameters at Line 18 using PPO-clip where $1+\epsilon$ occurs when the advantage estimation is positive else $1-\epsilon$ is used from Lines 19 to 23. The PPO-clip imposes the penalty on the $r(\phi)$ ratio in both cases. The fourth step is to update the critic parameter ($\theta_{new}$) at Line 24. 

The output to the algorithm is the actor-network that makes the decision to play the chunk by chunk with a specified bitrate at Line 29,  the critic network evaluates the state-value of the policy with PPO-clip for maximizing rewards  at Line 30 and the actor and critic parameters are updated based on the actor and the critic loss functions at Line 31. The PPO-ABR trains multiple agents in parallel, so the multi-agents are trained with their environments for each batch iteration. Moreover, the actor and critic parameters are updated using PPO-clip for each batch iteration. The value function parameters are updated after multiple epochs instead of a single epoch. Further, the central agent collects the mini-batch samples and updates the gradient to the next batch iterations. Overall, PPO-ABR results in a stable update and provides the bit rate to encode the next chunk.

\begin{algorithm}
\caption{PPO-ABR Algorithm}\label{alg:2}
\begin{algorithmic}[1]
\STATE \textbf{Input:} video samples, hyperparameters;
\STATE \textbf{Parameters:}
\STATE \textbf{Video vi;} {choose a video file as a input}
\STATE \textbf{Chunk c;} {select the bitrate for future chunks from video file}

\STATE \textbf{Initialize} the batch size $B$, clipping parameter $\epsilon$   
\STATE \textbf{Initialize} weight parameters: $\theta, \phi$ 
\FOR {video vi= 1,2,3...., VI}
\STATE Observe initial state $s_t$;
\FOR {chunk c=1,2,3...., C}
\STATE $\overline{V}_{\theta} = \sum_{k=1}^{K}V(s_{t};\theta_{k})$ for all states $s_{t}$
\STATE $R \leftarrow 0$ for terminal state $s_{terminal}$
\STATE $R = \overline{V}_{t}$ for non terminal states $s_{t}$
\FOR {each batch iteration}
    \STATE Compute advantage function on $B$
    \STATE  $A_\phi(s,a) = Q^\pi_\phi (s,a) - V^\pi_\phi (s)$
    \STATE Compute the importance sampling weight
    \STATE $r(\phi)= \frac{\pi_{\phi_{new}}(s,a)}{\pi_{\phi}(s,a)}$ using policy parameters
       \STATE Update actor parameter by PPO-clip:   
      $\text{maximize}_\phi  
    L^{clip}(\phi_{new}) =$ 
     \resizebox{0.85\hsize}{!}{
    $\kappa E_t\Bigg[\min\Bigg( L^{CPI}(\phi), clip(r(\phi) \,,1-\epsilon, 1+\epsilon)A_\phi(s,a)\Bigg)\Bigg]$ 
    }
    \IF{$A_\phi(s,a)\geq$0}
    \STATE clip$(r(\phi),1+\epsilon)A_\phi(s,a)$
    \ELSE
    \STATE  clip$(r(\phi),1-\epsilon)A_\phi(s,a)$ 
    \ENDIF   
    \STATE Update critic parameter    
     $\theta_{new}= \theta + \frac{\partial(R - \overline{V}_{\theta})^{2}}{\partial\theta}$ 
\ENDFOR
\ENDFOR
\ENDFOR
\STATE \textbf{Output:}
\STATE Actor network makes the decision to play the chunk by chunk with a specified bitrate
\STATE Critic network evaluates the state-value of the policy with PPO-clip for maximizing rewards  
\STATE Update actor and critic parameters $\theta,\phi$
\end{algorithmic}
\end{algorithm}

\section{Experimental details and Results}\label{section:Experimental details}
This section will describe the experimental methodology utilised for this study. This will include a description of the datasets used, the training method employed, the algorithms used for comparison, and the performance metrics used to assess their efficacy.

\subsection{Datasets}
\label{section:datasets}
We utilised multiple datasets FCC \cite{FCC}, Norway \cite{NORWAY}, LIVE \cite{LIVE}, OBOE \cite{OBOE}  for our experimentation, including both broadband and mobile datasets. First, we utilised the FCC  \cite{FCC} and Norway datasets \cite{NORWAY}, which include fixed broadband technologies and Telenor's 3G/HSDPA mobile wireless network. We utilized 59 and 68 traces from FCC and Norway throughput traces, respectively for our experiments. The range of throughput for both datasets is 0 to 6 Mbps. Secondly, we used live video streaming datasets \cite{LIVE}, which consists of data from wireless networks such as WiFi and 4G. The throughput range of these traces is between 0.2 Mbps and 4 Mbps, and 100 traces are utilised in our experiments. Lastly, we utilised OBOE dataset \cite{OBOE}, which include 428 traces from 500 video streaming sessions. Each OBOE trace stores the bandwidth measurements collected from wired, wireless, and cellular connections, and the throughput range is between 0 and 3 Mbps.



\subsection{Methodologies for Training, Comparative Algorithms, and Performance Metrics}

\begin{table}[h!]
\caption{Hyperparameters used during the training for Pensieve, SAC-ABR, and PPO-ABR.}
\centering
\resizebox{8cm}{!}{  
\begin{tabular}{|c|c|c|c|}
\hline
Hyperparameter              & Description            & Value    &Actor-critic algorithms                    \\ \hline
\(\gamma\) & Discount factor        & 0.99 & Pensieve, SAC-ABR, PPO-ABR    \\ \hline
\(\alpha_p\) & Actor network's  learning rate    & 0.0001 & Pensieve, SAC-ABR, PPO-ABR  \\ \hline
\(\alpha_c\)   & Critic network's  learning rate   & 0.001 & Pensieve, SAC-ABR, PPO-ABR   \\ \hline
\(\eta\)  & Entropy regularization  factor range  & 6 to 0.01 & Pensieve, SAC-ABR, PPO-ABR   \\ \hline
\(\tau\)   & Interpolation factor & 0.995  & SAC-ABR                        \\ \hline
$\epsilon$ &clipping parameter             & 0.2  &  PPO-ABR    \\ \hline
R &Random seed            & 42  &  PPO-ABR    \\ \hline

$n_{act}$             & Total number of agents & 16  & Pensieve, SAC-ABR, PPO-ABR    \\ \hline
\end{tabular}
}

\label{tab:hyperparameters}

\end{table}


We train PPO-ABR on the aforementioned datasets for 100,000 iterations, and then we choose the model with the highest average reward. Table \ref{tab:hyperparameters} summarizes the hyperparameters utilized for  PPO-ABR training. Specifically, clipped probability hyperparameter $\epsilon=0.2$ determines how much the new policy deviates from the old policy. These values have been selected based on the previous works \cite{pensieve}, \cite{OBOE}, and \cite{LIVE}. 
We use $n_{act} = 16$ agents for all our experiments. Finally, the performance of the proposed  PPO-ABR is compared to that of the following state-of-the-art DRL-based and non-DRL-based ABR algorithms: SAC-ABR \cite{sac-abr}, Pensieve \cite{pensieve}, BB \cite{bb}, RB \cite{rb}, BOLA \cite{bola}, and Robust-MPC \cite{mpc}. 








We compare the performance of all ABR algorithms using QoE \cite{sac-abr} as a metric. The QoE is expressed as:
\begin{equation}
    QoE = \sum_{n=1}^{N}q(b_{n}) - \mu \sum_{n=1}^{N}T_{n} - \sum_{n=1}^{N-1} \left| q(b_{n+1})-q(b_{n})\right|
\label{eq:qoe}
\end{equation}

The QoE is composed of three elements: (i) the total bit rates of all video chunks, (ii) the penalty incurred by re-buffering, and (iii) the video's smoothness, which is assessed by calculating the difference in bit rates used to encode consecutive chunks. Various versions of the QoE metric are examined in this context as follows: (i) \(QoE_{lin}\): \(q(b_{n}) = b_{n}\) with rebuffer penalty as \(\mu = 4.3\) and (ii) \(QoE_{log}\): \(q(b_{n})=log(b/b_{min})\) with \(\mu = 2.66\).

Note that we have utilized the above QoE metric formulation since it is commonly used in several other works including Robust-MPC \cite{mpc}, \cite{pensieve}, \cite{OBOE}, \cite{32}, \cite{stick} and \cite{sac-abr}. There also exist other QoE metric formulations, for example in \cite{refa} and \cite{refb}, that can also be used for the performance evaluation. In this work, we focus only on the QoE metric defined in Equation \ref{eq:qoe}.


\begin{figure}
\centering
\includegraphics[width=0.7\linewidth]{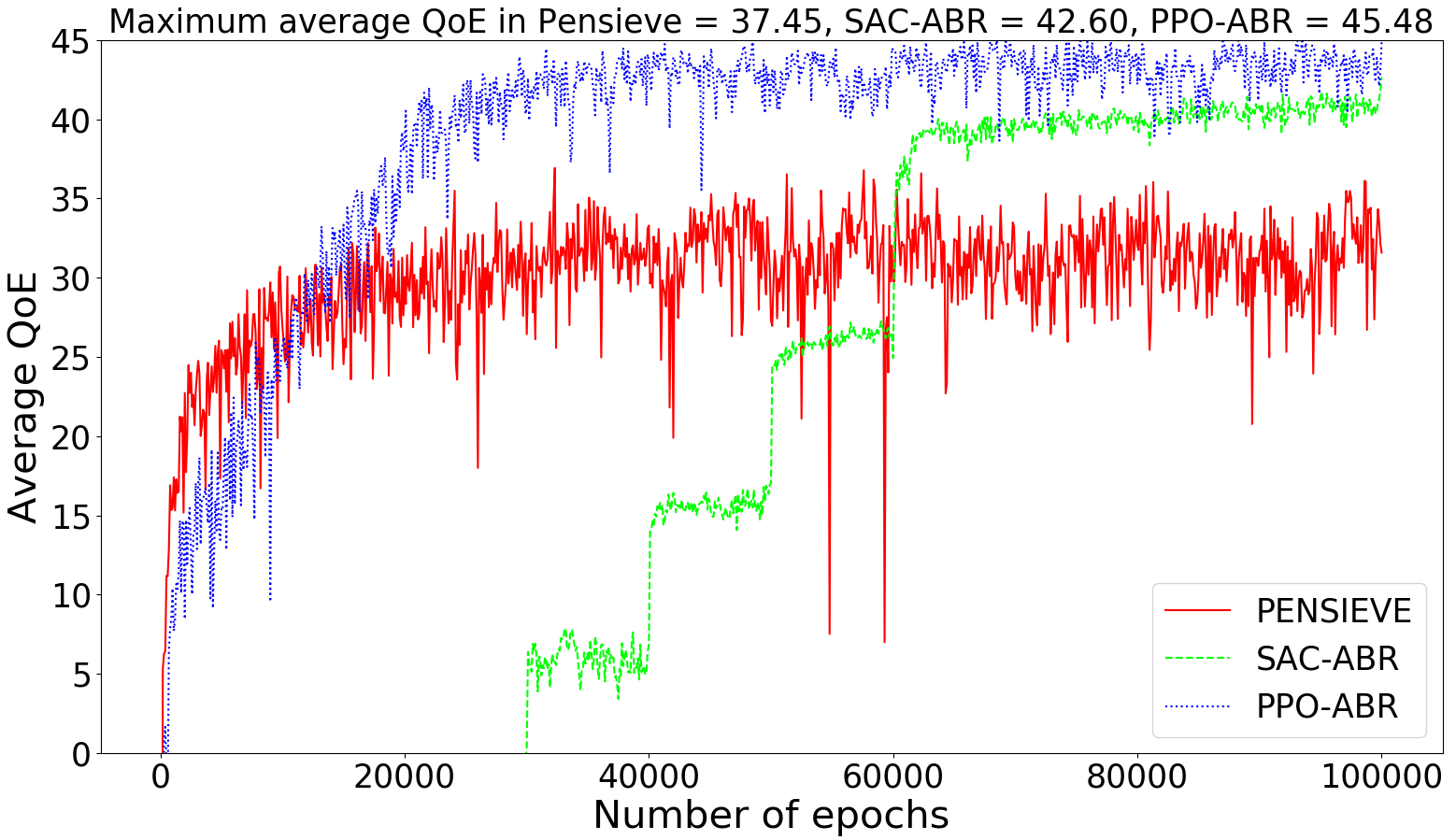}
\caption{The QoE performance of Pensieve, SAC-ABR, and PPO-ABR was measured during training over 100,000 epochs for the  $QoE_{lin}$ metric on FCC and Norway traces, and the average values were obtained.}
\label{fig:fcc_train}
\end{figure}

\begin{figure}
\centering
\includegraphics[width=0.7\linewidth]{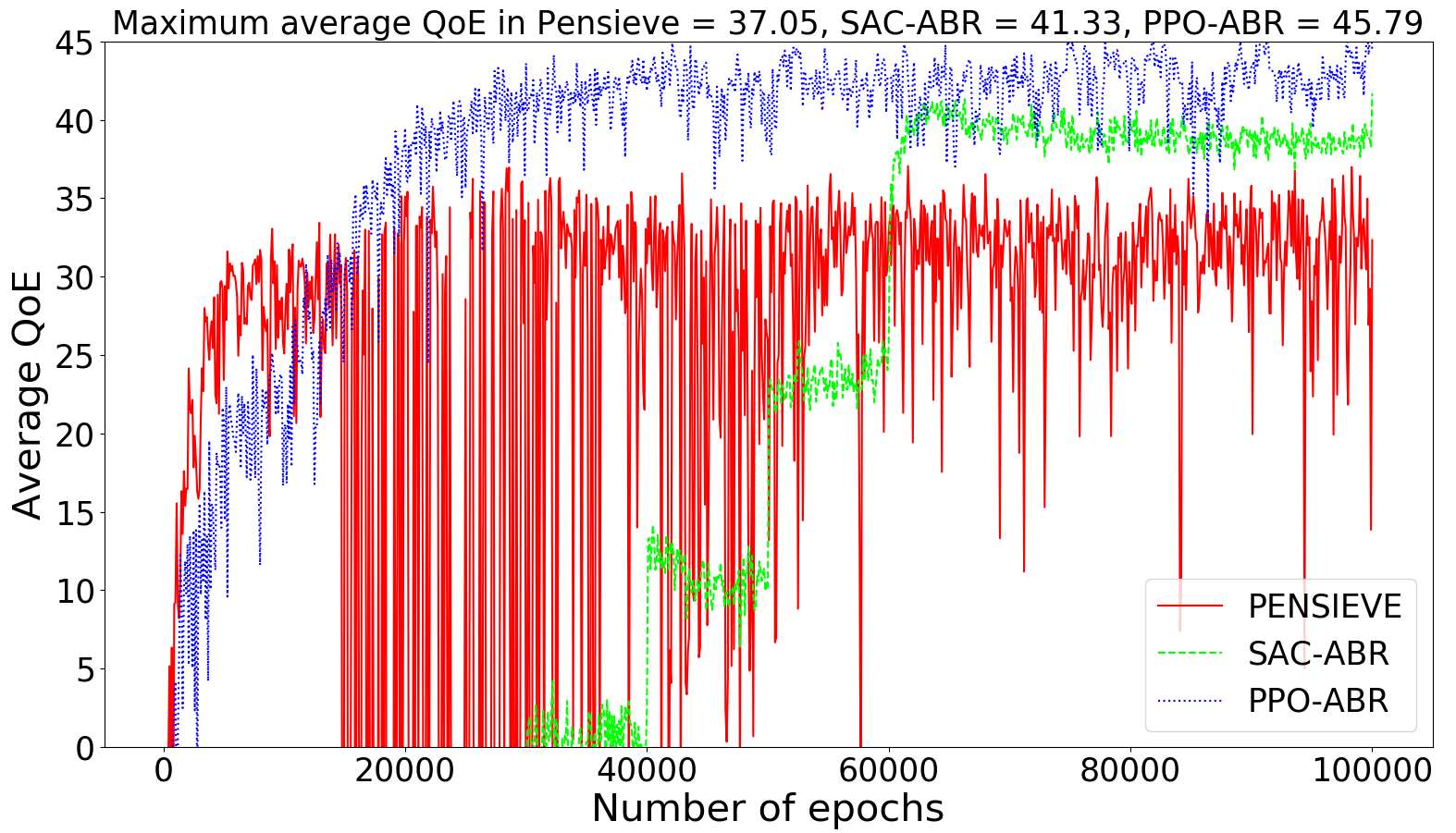}
\caption{The QoE performance of Pensieve, SAC-ABR, and PPO-ABR was measured during training over 100,000 epochs for the  $QoE_{lin}$ metric on OBOE traces, and the average values were obtained.}
\label{fig:oboe_train}
\end{figure}

\begin{figure}
\centering
\includegraphics[width=0.7\linewidth]{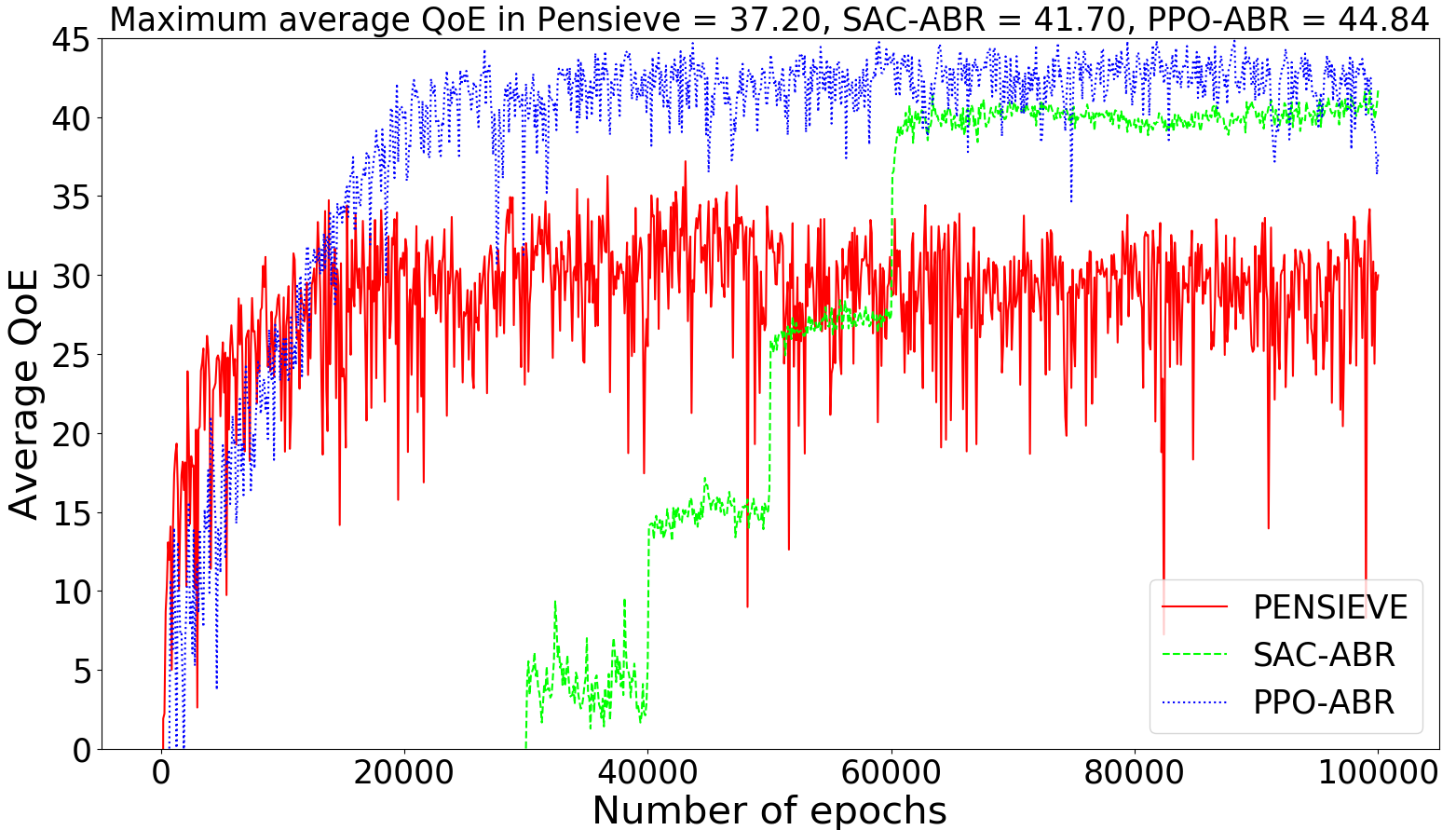}
\caption{The QoE performance of Pensieve, SAC-ABR, and PPO-ABR was measured during training over 100,000 epochs for the  $QoE_{lin}$ metric on Live traces, and the average values were obtained.}
\label{fig:live_train}
\end{figure}

\begin{table}[h!]
\caption{Training outcomes of Pensieve, SAC-ABR, and PPO-ABR concerning the $QoE_{lin}$ and $QoE_{log}$ metrics across multiple datasets.}
\centering
\resizebox{8cm}{!}{ 
\begin{tabular}{|c|c|c|c|c|c|c|}
\hline
\multirow2{*}{RL algorithm} & \multicolumn{2}{c|}{FCC Norway Traces}                                 & \multicolumn{2}{c|}{OBOE Traces} & \multicolumn{2}{c|}{Live traces} \\ \cline{2-7} 
                           & \multicolumn{1}{l|}{$QoE_{lin}$} & \multicolumn{1}{l|}{$QoE_{log}$} & \multicolumn{1}{l|}{$QoE_{lin}$} & \multicolumn{1}{l|}{$QoE_{log}$} & \multicolumn{1}{l|}{$QoE_{lin}$} & \multicolumn{1}{l|}{$QoE_{log}$} \\ \hline
\textbf{PPO-ABR}                  & \textbf{45.48}                      & \textbf{45.40}                       & \textbf{45.79}                        & \textbf{46.36}                       & \textbf{44.84}                        & \textbf{45.89}                       \\ \hline
SAC-ABR                  & 42.60                       & 45.20                      & 41.33                        & 43.88                       & 41.70                        & 43.46                       \\ \hline

Pensieve                  & 37.45                        & 37.84                       & 37.05                        & 36.30                       & 37.20                        & 37.59                       \\ \hline

\end{tabular}
}

\label{tab:testreslossbw1}
\end{table}

\subsection{Training results}

We trained PPO-ABR, SAC-ABR, and Pensieve using the three datasets mentioned in the preceding section.  Furthermore, in order to maximize entropy, we utilized an entropy regularization ranging from 6 to 0.01 for a better exploration-exploitation tradeoff, i.e., initially, an entropy value of six is used for a few iterations, and then it is gradually decreased to 0.01. It takes approximately eight hours to generate the training model for every algorithm with each dataset. 
Table \ref{tab:testreslossbw1} summarizes the QoE metrics obtained during training for the three datasets. The findings indicate that across all three datasets and for both $QoE_{lin}$ and $QoE_{log}$ metrics, PPO-ABR consistently outperforms SAC-ABR and Pensieve, achieving higher QoE metrics.


\begin{figure}
\centering
\includegraphics[width=0.8\linewidth]{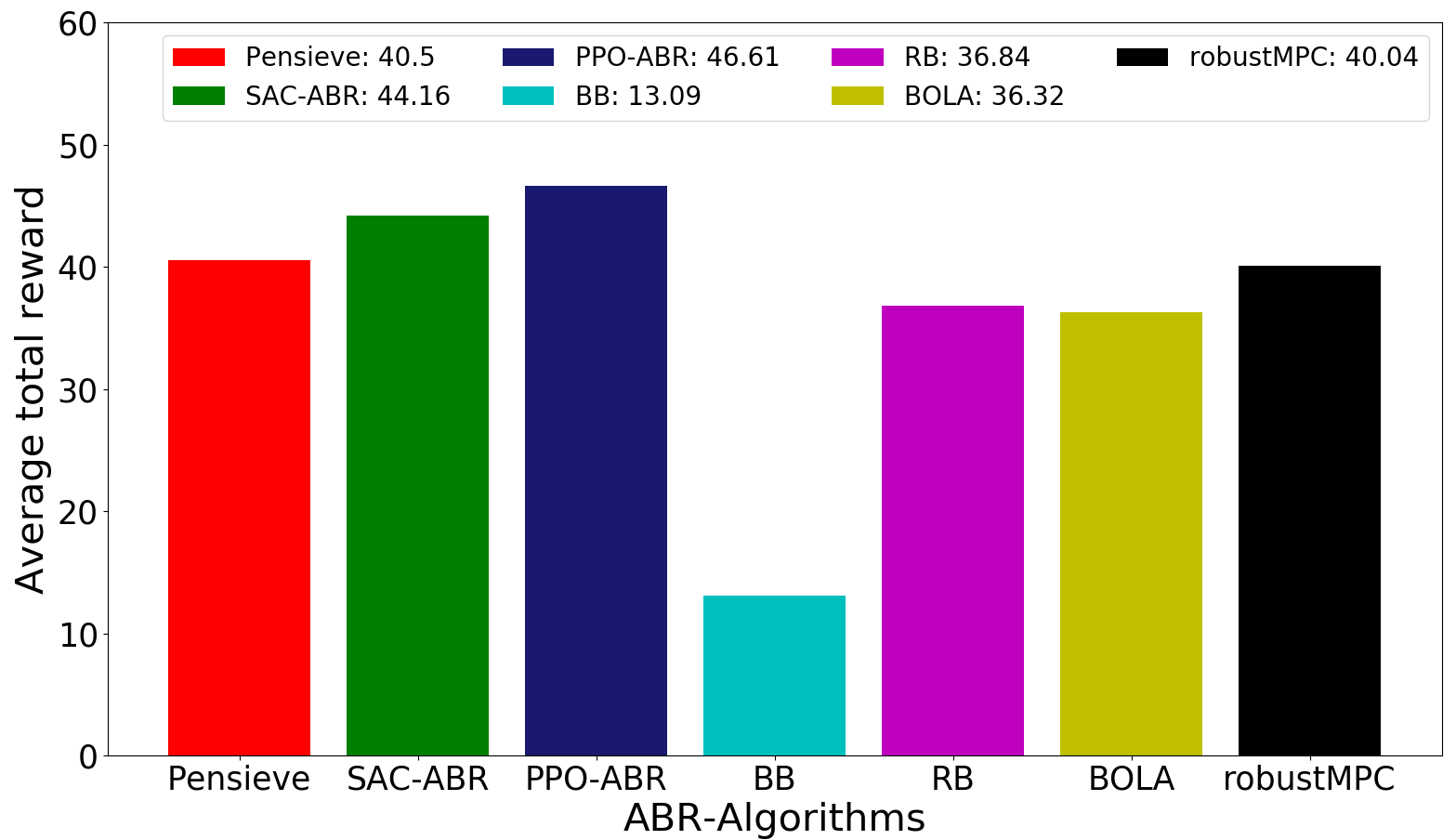}
\caption{Performance evaluation of ABR algorithms with $QoE_{lin}$ metric when tested on the model trained with FCC and Norway traces while the network is emulated with no packet loss.}
\label{fig:fig5}
\end{figure}

\begin{figure}
\centering
\includegraphics[width=\linewidth]{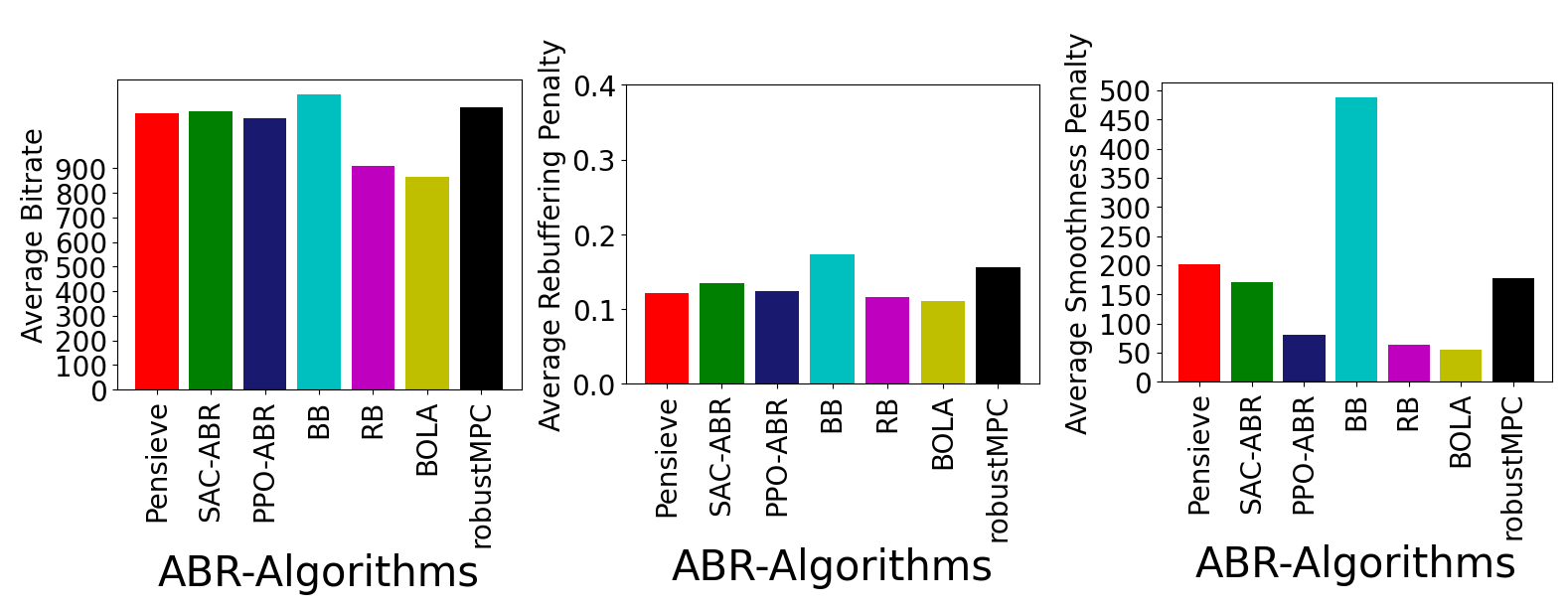}
\caption{Comparing PPO-ABR with current ABR methods by analyzing
their performance on the individual elements for $QoE_{lin}$ metric with no packet loss under emulation 
(Equation \ref{eq:qoe}).}
\label{fig:fig6}
\end{figure}

Figure \ref{fig:fcc_train} presents the average QoE value achieved by PPO-ABR, SAC-ABR, and Pensieve algorithms at each training epoch. We
can observe that SAC-ABR performs poorly at the initial stages of training due to high exploration. Our results show different behavior for each of these algorithms when the number of epochs increases during the training. The PPO-ABR achieves a high QoE value right from the start of the training. Similar improvements are observed with OBOE in Figure \ref{fig:oboe_train} and Live traces in Figure \ref{fig:live_train} as well, where Table \ref{tab:testreslossbw1} presents the values of QoE obtained using different ABR algorithms. 

\subsection{Testing results}

The training models are evaluated using the Mahimahi simulator \cite{mahimahi}. We used 250 traces from the Norway test datasets and 205 traces from the FCC test datasets to test the models, as stated in \cite{pensieve}. At a bit rate of 12 Mbps and a latency of 30 ms throughout the testing phase, we assessed how well each ABR algorithm performed. Figure \ref{fig:fig5} displays the average total reward obtained by various ABR algorithms with the $QoE_lin$ metric for each trace when the network is simulated during testing with no packet loss. According to our findings, the PPO-ABR algorithms have a higher average QoE of 46.61 than other ABR algorithms.

\begin{table}[h!]
\caption{On three datasets, the average QoE was attained using two different QoE metrics during simulation with no packet losses.}
\resizebox{8cm}{!}{ 
\centering
\begin{tabular}{|c|c|c|c|c|c|c|}
\hline
\multirow2{*}{ABR algorithm} & \multicolumn{2}{c|}{FCC and Norway traces}                                 & \multicolumn{2}{c|}{OBOE traces}                                 & \multicolumn{2}{c|}{Live traces}                                 \\ \cline{2-7} 
                           & \multicolumn{1}{l|}{$QoE_{lin}$} & \multicolumn{1}{l|}{$QoE_{log}$} & \multicolumn{1}{l|}{$QoE_{lin}$} & \multicolumn{1}{l|}{$QoE_{log}$} & \multicolumn{1}{l|}{$QoE_{lin}$} & \multicolumn{1}{l|}{$QoE_{log}$} \\ \hline
\textbf{PPO-ABR}                 & \textbf{46.61}                        & \textbf{44.93}                       & \textbf{45.09}                        & \textbf{46.25}                       & \textbf{46.91}                        & \textbf{45.68}                       \\ \hline

SAC-ABR                  & 42.77                        & 43.68                       & 39.72                        & 45.41                       & 42.59                        & 43.90                       \\ \hline

Pensieve                  & 39.63                        & 35.26                       & 37.96                        & 37.01                       & 39.12                        & 41.68                       \\ \hline

BB                  & 12.03                        & 12.78                       & 14.08                        & 20                      & 13.81                        & 20.26                      \\ \hline
RB                         & 35.62                       & 36.45                       & 36.22                        & 37.31                        & 37.45                        & 37.35                       \\ \hline
BOLA                         & 34.26                        & 35.30                       & 35.04                       & 37.09                      & 35.82                        & 36.05                       \\ \hline
Robust-MPC                       & 39.93                        & 40.44                       & 40.18                        & 38.29                       & 40.59                        & 38.99                      \\ \hline
\end{tabular}
}
\label{tab:table3}
\end{table}


In Figure \ref{fig:fig6}, we compare various ABR algorithms using the average playback bitrate, rebuffering penalty, and smoothness penalty for the $QoE_{lin}$ metric under emulation with no packet losses during testing in order to understand and illustrate the better performance of the PPO-ABR. Our findings indicate that, with the exception of BOLA and RB, most ABR algorithms attain greater bitrates. Several of these algorithms experience rebuffering penalties due to the higher bitrate choice, with BB and SAC-ABR having the biggest rebuffering penalties. Similarly, BB likewise has a significant smoothness penalty. The PPO-ABR delivers a higher average bit rate and, in comparison, lower smoothness and rebuffering penalties. The PPO-ABR achieves an average QoE higher than the other ABR algorithms due to the combined effects of these individual components.  The average QoE values attained by the ABR algorithms when evaluated on the network emulated with no packet losses are then shown in Table \ref{tab:table3}  for various QoE metrics.

\section{Conclusion}\label{section:Conclusion}



We have shown in this study the advantages of adopting on-policy DRL-based PPO-ABR to increase QoE for video streaming. Our suggested method specifically overcomes the limitations currently faced by state-of-the-art DRL-based methods and consistently achieves higher average QoE than SAC-ABR and Pensieve, respectively, by up to $13.52\%$ and $27.42\%$, and even higher QoE when compared to other conventional fixed-rule-based ABR algorithms. Future studies will examine PPO-ABR for edge-driven video distribution services and evaluate it using various QoE metric versions.

\section*{Acknowledgment}

This work has been supported by TCS foundation under the TCS research scholar program, 2019-2023, India.

\bibliographystyle{IEEEtran}
\bibliography{sample-base}

\end{document}